\documentclass[12pt,preprint]{aastex}
\usepackage{natbib}
\usepackage{graphicx}
\shorttitle{dark matter density spike around the supermassive black hole}
\begin{document}
\title{The first robust evidence showing a dark matter density spike around the supermassive black hole in OJ 287}
\author{Man Ho Chan \& Chak Man Lee}
\affil{Department of Science and Environmental Studies, The Education University of Hong Kong, Hong Kong, China}
\email{chanmh@eduhk.hk}

\begin{abstract}
Black hole dynamics suggests that dark matter would re-distribute near a supermassive black hole to form a density spike. However, no direct evidence of dark matter density spike around a supermassive black hole has been identified. In this letter, we present the first robust evidence showing a dark matter density spike around a supermassive black hole. We revisit the data of the well-known supermassive black hole binary OJ 287 and show that the inclusion of the dynamical friction due to a dark matter density spike around the supermassive black hole can satisfactorily account for the observed orbital decay rate. The derived spike index $\gamma_{\rm sp}=2.351^{+0.032}_{-0.045}$ gives an excellent agreement with the value $\gamma_{\rm sp}=2.333$ predicted by the benchmark model assuming an adiabatically growing supermassive black hole. This provides a strong verification of the canonical theory suggested two decades ago modeling the gravitational interaction between collisionless dark matter and supermassive black holes.
\end{abstract}


\section{Introduction}
In the past few decades, various studies showed that a supermassive black hole (SMBH) can alter the nearby dark matter density distribution to form a density spike \citep{Young,Gondolo,Merritt,Gnedin,Merritt2,Sadeghian,Nampalliwar}. The dark matter density would be steepened within the sphere of influence of a SMBH due to the conservation of angular momentum and radial action. Consequently, we expect that such a high dark matter density near a SMBH would significantly trigger the rate of dark matter annihilation to give a strong emission of high-energy gamma-rays \citep{Bertone,Gnedin,Fields,Shapiro}. However, no strong gamma-ray emission has been detected so far near any SMBH, including the SMBH in our Galaxy (Sgr A*) \citep{Fields}. 

Recently, \citet{Chan} claim that the data of two nearby black hole low-mass X-ray binaries, A0620-00 and XTE J1118+480, might reveal the existence of dark matter density spikes around their black holes. The dynamical friction exerted by the dark matter density spikes can satisfactorily explain the abnormal large orbital decay rates of the companion stars in these two binaries. However, the evidence of any dark matter density spike around a SMBH is definitely lacking, even though many recent studies are still modelling the black hole binary inspirals with the existence of dark matter density spikes \citep{Yue,Tang,Dai,Becker,Li,Qunbar}. In particular, we expect that the gravitational waves (GWs) emitted by supermassive black hole binaries (SMBHBs) can help reveal the properties of SMBHs and verify our theoretical understanding about these exotic systems, although the emission of these low-frequency GWs can only be detected by future space GW interferometers. Moreover, confirming the existence of a dark matter density spike can also help constrain the parameters of dark matter annihilation or the decay rate \citep{Gondolo,Kar}. 

In this letter, we present the first evidence of the existence of a dark matter density spike around a SMBH. We revisit the data of the well-known SMBHB OJ 287, which consists of a secondary SMBH with mass $m_{\rm BH} \approx 1.5 \times 10^8M_{\odot}$ orbiting a primary SMBH with mass $M_{\rm BH} \approx 1.8 \times 10^{10}M_{\odot}$. We show that the energy loss rate due to GW emission is significantly smaller than the observed energy loss rate, with a discrepancy of more than $4.3 \sigma$. With the assumption of a dark matter density spike around the primary SMBH, the dynamical friction can provide the extra energy loss rate which can satisfactorily account for the observed energy loss rate. The derived spike index gives an excellent agreement with the predicted value based on the adiabatically growing SMBH model \citep{Gondolo,Fields}. Therefore, this is the strongest evidence so far to reveal the existence of dark matter density spike around a SMBH. 

\section{The supermassive black hole binary OJ 287}
OJ 287 is a well-known SMBHB as it has been studied for more than a century \citep{Sillanpaa,Valtonen,Valtonen4,Valtonen2,Dey,Dey2,Laine,Komossa,Titarchuk,Valtonen3,Zwick,Martinez}. This SMBHB contains one secondary SMBH orbiting another very massive primary SMBH so that it is one of the most exotic systems observed in our universe. This binary emits a large amount of X-ray and radio radiation and we expect that it could also emit a huge amount of GW energy. Based on the analysis of the accurately extracted starting epochs of ten optical outbursts of OJ 287 between the years of 1912-2016, an accurate orbit of the secondary SMBH in OJ 287 can be determined \citep{Dey}. This study has provided the most robust results and obtained accurate orbital parameters with very small uncertainties for OJ 287 (see Table 1 for the essential orbital parameters).  

Although there are some recent studies claiming that the mass of the primary SMBH is only $M_{\rm BH} \sim 10^8M_{\odot}$ \citep{Komossa} and the mass of the secondary SMBH is smaller by 20\% \citep{Titarchuk}, these results are respectively based on the analysis on a particular outburst observation in 2022 \citep{Komossa} and the comparative study focusing on X-ray data only \citep{Titarchuk}. Overall speaking, the orbital parameters given in \citet{Dey}, which have followed the outburst data in the past 104 years, are still the most comprehensive and robust results for OJ 287. In the followings, we will base on these robust orbital parameters shown in Table 1 to perform our analysis. 

As orbital precession exists in OJ 287, the orbit of the secondary SMBH can be described by the relative distance $r(\phi)$ between the primary and secondary SMBHs:
\begin{equation}
r(\phi)=\frac{a(1-e^2)}{1+e \cos(1-\alpha)\phi},
\end{equation}
where $a$ is the orbital semi-major axis, $e$ is the eccentricity, and $\alpha=\Delta \Phi/2\pi$ is the precession phase angle. 

One of the most intriguing properties of OJ 287 is that the orbit of the secondary SMBH is shrinking. The orbital period decay rate is $\dot{P}=-(0.00099 \pm 0.00006)$ \citep{Dey}, which means that two SMBHs will merge together after about 12000 years. The expected reason for this orbital shrinking is that energy is given out continuously due to GW emission \citep{Valtonen4,Dey}. Previous studies following pulsar binaries have shown that the orbital shrinking rate of pulsars agrees with the predicted rate based on GW emission \citep{Weisberg}. Based on General Relativity, the energy loss rate due to GW emission can be analytically given by \citep{Peters,Maggiore,Tang,Li}
\begin{equation}
\dot{E}_{\rm GW}=-\frac{32G^4\mu^2M^3}{5c^5a^5}(1-e^2)^{-7/2}\left(1+\frac{73}{24}e^2+\frac{37}{96}e^4\right),
\label{dEdtGW}
\end{equation}
where $\mu=m_{\rm BH}M_{\rm BH}/(m_{\rm BH}+M_{\rm BH})$ and $M=m_{\rm BH}+M_{\rm BH}$ are the reduced mass and total mass of SMBHB OJ 287 respectively.

On the other hand, one can convert the orbital period decay rate to the total energy loss rate theoretically. Although the secondary SMBH is orbiting with a large precession angle, we can still apply the Keplerian relation $P \propto a^{3/2}$, though the proportionality constant is larger due to the precession effect. Also, since the General Relativistic terms only contribute about 1\% of the total mechanical energy $E$ and the orbital period decay rate is smaller than 0.1\%, the total mechanical energy of the orbital motion can be well-approximated by the conventional Newtonian expression $E=-GM\mu/2a$. Therefore, we can write the total energy loss rate in terms of the orbital period decay rate:
\begin{equation}
\dot{E}=-\frac{2E\dot{P}}{3P}.
\end{equation}

By using Eqs.~(2) and (3) with $M_{\rm BH}=(1.8348 \pm 0.0008)\times 10^{10}M_{\odot}$, $m_{\rm BH}=(1.5013 \pm 0.0025)\times 10^8M_{\odot}$, $P=12.067\pm 0.007$ years, $\dot{P}=0.00099 \pm 0.00006$, and $e=0.657 \pm 0.001$ obtained in \citet{Dey}, and taking $a=1.72 \times 10^{17}$ cm based on orbital analysis \citep{Valtonen,Laine,Martinez}, we get $\dot{E}=-(3.66 \pm 0.24) \times 10^{41}$ W and $\dot{E}_{\rm GW}=-(2.62 \pm 0.02) \times 10^{41}$ W. This gives a $4.3 \sigma$ discrepancy between $\dot{E}$ and $\dot{E}_{\rm GW}$, which shows a large tension between observations and theoretical prediction assuming solely GW emission. Therefore, we expect that there must exist another important energy loss mechanism in OJ 287.

\section{Dynamical friction of the dark matter density spike}
Based on the results of numerical simulations, the density profile of a massive halo formed by collisionless dark matter would follow the Navarro-Frenk-White (NFW) density profile \citep{Navarro}:
\begin{equation}
\rho_{\rm DM}=\frac{\rho_sr_s^3}{r(r+r_s)^2},
\end{equation}
where $\rho_s$ and $r_s$ are scale density and scale radius respectively. This profile is commonly modeled as the galactic dark matter density profile, including our Galaxy and the M31 galaxy \citep{Sofue}. Nevertheless, the galactic central SMBH would re-distribute dark matter to form a dark matter density spike around the SMBH due to conservation of angular momentum and radial action \citep{Gondolo}. Outside the spike region $r \ge r_{\rm sp}$, the dark matter density would follow back to the global NFW density profile. To summarize, the dark matter density around the primary SMBH can be described by the following spike model (with General Relativistic correction) \citep{Sadeghian,Eda,Tang,Capozziello,John}:
\begin{equation}
\rho_{\rm DM}=\left\{
\begin{array}{ll}
0 & {\rm for }\,\,\, r\le 2R_s \\
\rho_{\rm sp} \left(1-\frac{2R_s}{r} \right)^3 \left(\frac{r}{r_{\rm sp}} \right)^{-\gamma_{\rm sp}} & {\rm for }\,\,\, 2R_s <r \le r_{\rm sp}, \\
\frac{\rho_sr_s}{r} & {\rm for}\,\,\, r_{\rm sp}<r \ll r_s \\
\end{array}
\right.
\end{equation}
where $R_s=2GM_{\rm BH}/c^2$.

For the benchmark model suggesting an adiabatic growth of SMBH, one can relate the spike index $\gamma_{\rm sp}$ with the power-law index of the dark matter density outside the spike $\gamma$: $\gamma_{\rm sp}=(9-2\gamma)/(4-\gamma)$ \citep{Gondolo,Fields,Eda}. As the NFW profile suggests $\gamma=1$ for $r \ll r_s$, the adiabatic growth model predicts $\gamma_{\rm sp}=7/3 \approx 2.333$, which gives a very high dark matter density near a SMBH. Note that we did not include the effect of dark matter annihilation in our analysis. A large rate of dark matter annihilation would reduce the dark matter density spike to the so-called annihilation plateau density \citep{Fields}. As the age of the SMBH, mass of dark matter particles, and the annihilation cross section are unknown, we neglect the annihilation effect in our analysis. 

In the followings, we describe a theoretical framework to model the unknown parameters $\rho_s$, $r_s$, $\rho_{\rm sp}$ and $r_{\rm sp}$ by the known parameters $M_{\rm BH}$ and $m_{\rm BH}$. First of all, by considering 43 galaxy-scale strong gravitational lenses, there is an empirical relation between the mass of galactic SMBHs and the total dynamical mass of galaxies \citep{Bandara}:
\begin{eqnarray}
\log_{10}(M_{\rm BH}/M_{\odot})&=&(8.18\pm0.11)+(1.55\pm0.31)\nonumber\\
&&\times[\log_{10}(M_{\rm tot}/M_{\odot})-13.0].
\label{Mtot}
\end{eqnarray}
Surprisingly, this empirical relation gives an excellent agreement with the later simulation result $M_{\rm BH} \propto M_{\rm tot}^{1.55 \pm 0.05}$ \citep{Booth} and it is consistent with the results for elliptical galaxies $M_{\rm BH} \propto M_{\rm tot}^{1.6}$ \citep{Bogdan}. Based on the empirical relation in Eq.~(6) with the corresponding uncertainties, we get $M_{\rm tot}=2.21^{+3.67}_{-1.06} \times 10^{14}M_{\odot}$ for the galaxy hosting OJ 287. By taking this total dynamical mass as the virial mass, we can calculate the virial radius by $r_{200}^3=3M_{\rm tot}/4 \pi \rho_{200}$. Here, $\rho_{200}$ is defined as $\rho_{200}=200\rho_c$ with 
\begin{equation}
\rho_c = \frac{3H_0^2}{8 \pi G}[\Omega_{\rm m}(1+z)^3+\Omega_{\Lambda}+(1-\Omega_{\rm m}-\Omega_{\Lambda})(1+z)^2].
\end{equation}
By adopting the values of the redshift of OJ 287 $z=0.306$ \citep{Benitez}, the cosmological density parameters $\Omega_m=0.315 \pm 0.007$ and $\Omega_{\Lambda}=0.685 \pm 0.007$, and the Hubble constant $H_0=67.4\pm 0.5$ km/s/Mpc from Planck's observation \citep{Planck}, we get $r_{200}=3.54^{+1.36}_{-0.69} \times 10^{24}$ cm for the galaxy hosting OJ 287. 

Furthermore, using the mass-concentration relation of cosmological structures, we can get the concentration parameter $c_{200}$ from the total dynamical mass $M_{\rm tot}$. The empirical mass-concentration relation based on the lensing data of galaxies and galaxy clusters can be written as \citep{Xu}
\begin{equation}
c_{200}=C_0\left(\frac{M_{\rm tot}}{10^{12}M_{\odot}h^{-1}}\right)^{-\gamma_c}
\left[1+\left(\frac{M_{\rm tot}}{M_0}\right)^{0.4}\right],
\end{equation}
where $h=0.674$, $C_0=5.119^{+0.183}_{-0.185}$, $\gamma_c=0.205^{+0.010}_{-0.010}$ and $\log_{10}(M_0)=14.083^{+0.130}_{-0.133}$.

Again, by including the corresponding uncertainties, we get $c_{200}=4.17^{+0.99}_{-0.55}$ for OJ 287, which gives $r_s=r_{200}/c_{200}=8.49^{+1.01}_{-0.62} \times 10^{23}$ cm. Furthermore, since we know $M_{\rm tot}$ and $r_s$, we can get $\rho_s=6.84^{+4.23}_{-1.83} \times 10^{-26}$ g cm$^{-3}$. 

Based on the standard spike model, the spike radius $r_{\rm sp}$ is empirically defined by $r_{\rm sp}=0.2r_{\rm in}$, where $r_{\rm in}$ is the radius of influence \citep{Fields,Eda,Kavanagh}. The radius of influence can be determined by \citep{Merritt,Merritt2,Eda,Kavanagh}:
\begin{equation}
M_{\rm DM}(r \le r_{\rm in})= \int_0^{r_{\rm in}} 4\pi r^2 \rho_{\rm DM}dr=2M_{\rm BH}.
\end{equation}
Therefore, we get the following analytic relation \citep{Eda,Kavanagh,Mukherjee}
\begin{equation}
r_{\rm sp}=\left[\frac{(3-\gamma_{\rm sp})0.2^{3-\gamma_{\rm sp}}M_{\rm BH}}{2\pi\rho_{\rm sp}}\right]^{1/3}.
\end{equation}
Also, using Eq.~(5) and considering at $r=r_{\rm sp}$, we have 
\begin{equation}
\rho_sr_s=\rho_{\rm sp}r_{\rm sp} \left(1-\frac{2R_s}{r_{\rm sp}} \right)^3.
\end{equation}
The above two relations Eq.~(10) and Eq.~(11) can connect $\rho_{\rm sp}$ and $r_{\rm sp}$ individually with the spike index $\gamma_{\rm sp}$. 

Since the dark matter density is extremely high near the primary SMBH, the effect of dynamical friction would also be very large. In fact, the effect of dynamical friction of a dark matter density spike has been theorized for a long time. Most theoretical studies have included the effect of dynamical friction in modelling black hole mergers \citep{Dai,Becker,Qunbar,Mukherjee}. The energy loss rate due to dynamical friction is given by \citep{Chandrasekhar,Yue}
\begin{equation}
\dot{E}_{\rm DF}=-\frac{4\pi G^2 \mu^2 \rho_{\rm DM} \xi(\sigma) \ln \Lambda}{v},
\end{equation}
where $\ln \Lambda \approx \ln \sqrt{M_{\rm BH}/m_{\rm BH}}$ is the Coulomb Logarithm \citep{Kavanagh}, $\xi(\sigma) \approx 1$ is a numerical factor depending on the dark matter velocity dispersion $\sigma$, and $v=(GM_{\rm BH}/p)^{1/2} \sqrt{(e^2-1)+2[1+e \cos (1-\alpha)\phi]}$ is the orbital velocity, with $p=a(1-e^2)$ \citep{Tang}. If dark matter particles follow a Maxwellian distribution, the numerical factor can be described by $\xi(\sigma)={\rm erf}(X)-2Xe^{-X^2}/\sqrt{\pi}$, where $X=v/\sqrt{2}\sigma$ \citep{Merritt3}. Assuming the dark matter velocity dispersion is close to the velocity dispersion in the galactic bulge, we can get $\sigma \sim 440$ km/s by using the SMBH-velocity dispersion relation $M_{\rm BH}=10^{8.32\pm 0.05}M_{\odot}(\sigma/{\rm 200~km/s})^{5.64\pm 0.32}$ \citep{McConnell}. Since $v \sim (GM_{\rm BH}/p)^{1/2} \sim 50000$ km/s, we can get $X \sim 80$, which gives $\xi(\sigma)$ almost equal to 1. Hence, by using the dark matter density spike expression, we can get the average energy loss rate for one period due to dynamical friction (including precession effect):
\begin{eqnarray}
\dot{E}_{\rm DF}&=&-2G^{\frac32}\mu^2\rho_{\rm sp}r_{\rm sp}^{\gamma_{\rm sp}}(1-e^2)^{\frac32}\ln\Lambda \nonumber\\
&&\times \int_0^{2\pi} \frac{[1+e\cos(1-\alpha)\phi]^{\gamma_{\rm sp}-2}[p-2R_s(1+e\cos(1-\alpha)\phi)]^3}
{p^{\gamma_{\rm sp}+\frac52}M_{\rm BH}^{\frac12}[1+2e\cos(1-\alpha)\phi+e^2]^{\frac12}} d \phi.
\end{eqnarray}

As we mentioned that $\rho_{\rm sp}$ and $r_{\rm sp}$ depend on the spike index, from the above equation, we can see that the average energy loss rate due to dynamical friction depends on the spike index $\gamma_{\rm sp}$ only. By writing the total energy decay rate $\dot{E}=\dot{E}_{\rm GW}+\dot{E}_{\rm DF}$, we can constrain the value of the spike index $\gamma_{\rm sp}$. Including all of the uncertainties of the parameters and empirical relations, we can obtain the range of $\gamma_{\rm sp}$ by adopting $\dot{E}=-(3.66 \pm 0.24) \times 10^{41}$ W. As $\dot{E}_{\rm DF}$ depends sensitively on $\gamma_{\rm sp}$, from Fig.~1, we get a very narrow range of $\gamma_{\rm sp}=2.351^{+0.032}_{-0.045}$, which gives an excellent agreement with the canonical model prediction $\gamma_{\rm sp}=2.333$. This means that the inclusion of the dynamical friction due to dark matter density spike can satisfactorily account for the discrepancy in the energy loss rate. Note that stellar heating effect near the primary SMBH might drive the spike index down to a smaller value \citep{Gnedin,Merritt2}. Nevertheless, if the central total stellar mass near the primary SMBH of OJ 287 is $\sim 0.1$\% of $M_{\rm BH}$, the heating time required to change the spike index would be longer than 17 Gyr \citep{Merritt2}. Therefore, the stellar heating effect might not be significant for OJ 287. 

\section{Discussion}
In this letter, we have shown that the observed energy loss rate (i.e. the period decay rate) of OJ 287 is much larger than the predicted energy loss rate solely due to GW emission, with a discrepancy of $4.3 \sigma$. Nevertheless, by adding the energy loss rate of the dynamical friction due to the dark matter density spike around the primary SMBH, it can satisfactorily account for the discrepancy and reproduce the observed energy loss rate, with a narrow range of the spike index $\gamma_{\rm sp}=2.351^{+0.032}_{-0.045}$. Surprisingly, this spike index gives an excellent agreement with the canonical model prediction $\gamma_{\rm sp}=2.333$ based on the adiabatic SMBH growing model and the standard global NFW dark matter distribution predicted from numerical simulations. These provide a consistent picture to describe the period decay rate of OJ 287 and reveal the first robust evidence of the existence of a dark matter density spike around a SMBH. On the other hand, \citet{Alachkar} recently followed the orbital dynamics to constrain the dark matter spike mass of OJ 287. Our result shows that the dark matter spike mass is about 0.1\% of the primary SMBH mass, which is consistent with the constraints obtained in \citet{Alachkar} ($<3$\% of the primary SMBH mass).

Note that the spike index relation $\gamma_{\rm sp}=(9-2\gamma)/(4-\gamma)$ is derived based on the adiabatic growth model of a single SMBH \citep{Gondolo}. For SMBHBs like OJ 287, such a relation might not be applicable. Therefore, the agreement between our constrained $\gamma_{\rm sp}$ and the canonical value $2.333$ may be just a coincidence only. According to numerical simulation results, SMBHBs can scatter dark matter particles and decrease the density in the inner regions \citep{Merritt4,Merritt3}. Therefore, the spike index for OJ 287 would be smaller than the expected value. However, there are some mechanisms which can replenish the inner regions with dark matter particles and stars. For example, \citet{Zhao2} show that the efficient randomization of the orbits can provide a replenishment of the dark matter loss near the SMBHs. Also, \citet{Beraldo} show that the orbits of dark matter particles and stars could be destabilized and brought to the inner galactic region with the crossing of the main bar resonances. These mechanisms provide some possibilities for replenishing dark matter particles to the inner region of OJ 287. Therefore, the large $\gamma_{\rm sp}$ constrained in our study might still be possible. Hence, our result may provide an important clue to understand the complicated interactions and feedbacks between dark matter and SMBHBs.

In fact, \citet{Chan} have shown the possible existence of dark matter density spikes around stellar-mass black holes. Our results here might support the idea of theoretical prediction that dark matter density spikes might exist around most of the black holes, including intermediate-mass black holes \citep{Lacroix,Chan3} and SMBHs \citep{Gondolo}. In particular, the galactic SMBHs could be good targets for us to study the properties of dark matter, such as the annihilation and decay constraints. Besides, taking the SMBH in our galaxy as an example, dynamical studies using the stars orbiting the SMBH (Sgr A*) can be another way to examine the existence of any density spike around the SMBH \citep{Chan2,John,Shen}. Future accurate observations of the stars orbiting the SMBH can help verify the dark matter density spike model and constrain the properties of dark matter. 

Moreover, future low-frequency GW observations in space can further examine OJ 287 and other similar SMBHBs in our universe. The pattern of the GW signals can reveal the structure of SMBHBs and the SMBHB inspiral process \citep{Hannuksela,Zhao}. The low-frequency GW data can provide the final smoking-gun evidence to verify our result and our understanding of the interactions between dark matter and SMBHs. 

\begin{table}
\caption{Parameters of OJ 287 \citep{Dey}}
\begin{tabular}{|l||c|}
\hline\hline
Parameter& Value \\
\hline
Mass of the primary SMBH $M_{\rm BH}$ $(10^6M_{\odot})$ & $18348\pm 8$\\
Mass of the secondary SMBH $m_{\rm BH}$ $(10^6M_{\odot})$ & $150.13\pm 0.25$\\
Eccentricity $e$ & $0.657 \pm 0.001$\\
Precession angle per period $\Delta \Phi$ (deg) & $38.62\pm  0.01$\\
Orbital period $P$ (year) & $12.067 \pm 0.007$\\
Orbital period decay rate $\dot{P}$ & $-0.00099 \pm 0.00006$\\
\hline\hline
\end{tabular}
\label{table1}
\end{table}

\begin{figure}
\vskip 10mm
\includegraphics[width=140mm]{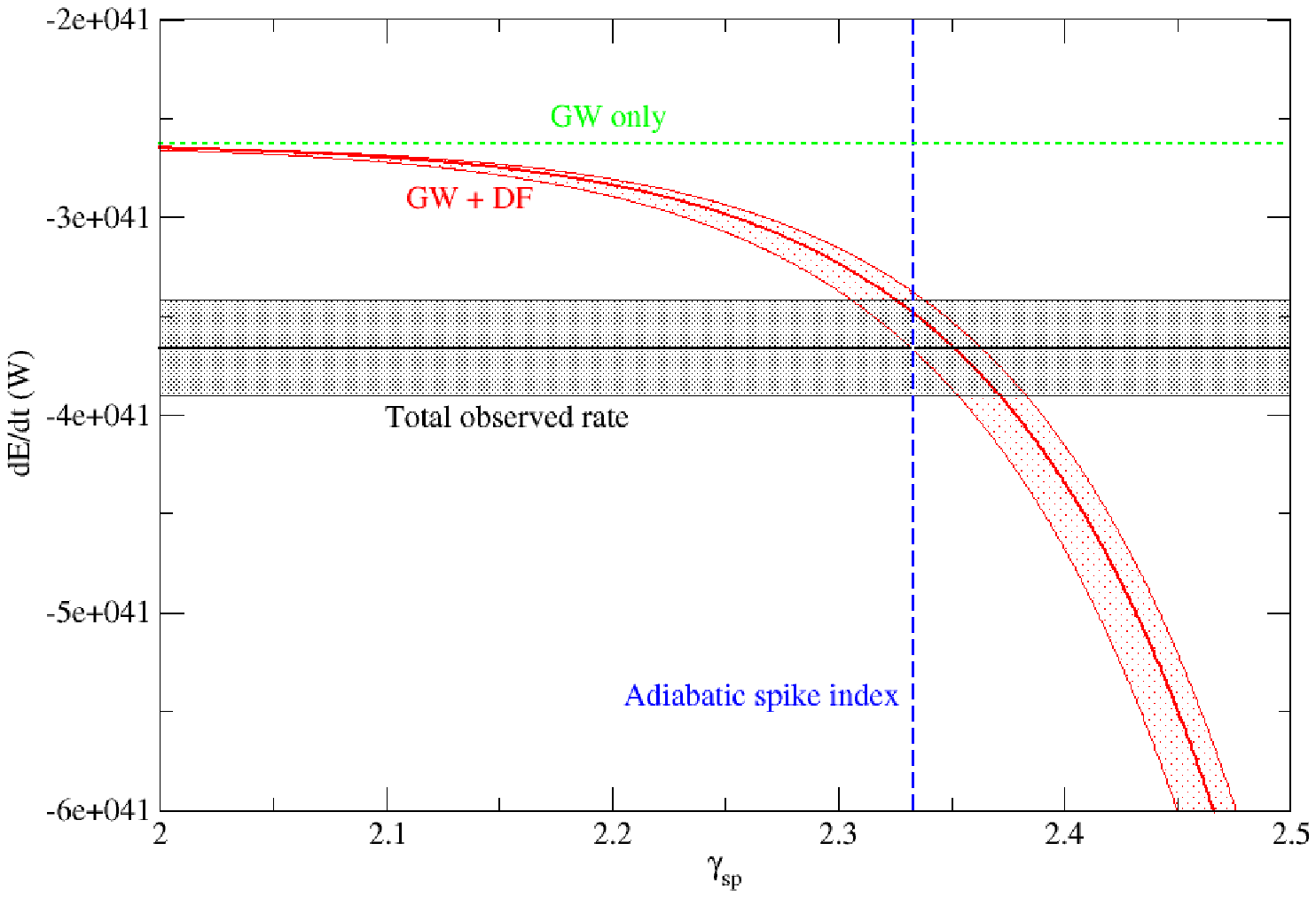}
\caption{The shaded region bounded by the black lines is the total energy decay rate constrained by observations (the thick black line indicates the average value). The shaded region bounded by the red lines is the predicted energy decay rate combining the effects of GW and dynamical friction (DF) of the dark matter density spike (the thick red line indicates the average value). The green dotted line indicates the energy decay rate due to GW emission only. The blue vertical dashed line indicates the spike index $\gamma_{\rm sp}=2.333$ predicted by the adiabatic SMBH growing model. Dynamical friction in a dark matter density spike with spike index $\gamma_{\rm sp}=2.351^{+0.032}_{-0.045}$ can account for the large orbital period decay rate measured by \citet{Dey}.}
\label{Fig1}
\vskip 5mm
\end{figure}

\section{Acknowledgements}
We thank the anonymous referee for useful comments. The work described in this paper was partially supported by a grant from the Research Grants Council of the Hong Kong Special Administrative Region, China (Project No. EdUHK 18300922).


\begin{thebibliography}{}
\bibitem[Alachkar, Ellis \& Fairbairn (2023)]{Alachkar} Alachkar A., Ellis J. \& Fairbairn M., 2023, Phys. Rev. D 107, 103033.
\bibitem[Bandara, Crampton \& Simard (2009)]{Bandara} Bandara K., Crampton D. \& Simard L., 2009, Astrophys. J. 704, 1135.
\bibitem[Becker et al. (2022)]{Becker} Becker, N., Sagunski, L., Prinz, L., \& Rastgoo, S., 2022,  Phys. Rev. D 105, 063029.
\bibitem[Benitez \& Dultzin-Hacyan (1996)]{Benitez} Benitez E. \& Dultzin-Hacyan D., 1996, Astrophys. J. 464, L47.
\bibitem[Beraldo e Silva et al. (2023)]{Beraldo} Beraldo e Silva L., Debattista V. P., Anderson S. R., Valluri M., Erwin P., Daniel K. J. \& Deg N., 2023, Astrophys. J. 955, 38.
\bibitem[Bertone et al. (2002)]{Bertone} Bertone, G., Sigl, G. \& Silk, J., 2002, Mon. Not. R. Astron. Soc. 337, 98.
\bibitem[Bogd\'an \& Goulding (2015)]{Bogdan} Bogd\'an A. \& Goulding A. D., 2015, Astrophys. J. 800, 124.
\bibitem[Booth \& Schaye (2010)]{Booth} Booth C. M. \& Schaye J., 2010, Mon. Not. R. Astron. Soc. 405, L1.
\bibitem[Capozziello, Zare \& Hassanabadi (2023)]{Capozziello} Capozziello S., Zare S. \& Hassanabadi H., 2023, arXiv:2311.12896.
\bibitem[Chan (2018)]{Chan3} Chan M. H., 2018, Mon. Not. R. Astron. Soc. 481, 3618.
\bibitem[Chan, Lee \& Yu (2022)]{Chan2} Chan M. H., Lee C. M. \& Yu C. W., 2022, Sci. Rep. 12, 15258.
\bibitem[Chan \& Lee (2023)]{Chan} Chan M. H. \& Lee C. M., 2023, Astrophys. J. 943, L11.
\bibitem[Chandrasekhar (1943)]{Chandrasekhar} Chandrasekhar S., 1943, Astrophys. J. 97, 255.
\bibitem[Dai et al. (2022)]{Dai} Dai, N., Gong, Y., Jiang, T. \& Liang, D. 2022, Phys. Rev. D 106, 064003.
\bibitem[Dey et al. (2018)]{Dey} Dey L. {\it et al.}, 2018, Astrophys. J. 866, 11.
\bibitem[Dey et al. (2019)]{Dey2} Dey L. {\it et al.}, 2019, Universe 5, 108.
\bibitem[Eda et al. (2015)]{Eda} Eda, K., Itoh, Y., Kuroyanagi, S. \& Silk, J., 2015, Phys. Rev. D 91, 044045.
\bibitem[Fields et al. (2014)]{Fields} Fields, B. D. , Shapiro, S. L. \& Shelto, J., 2014, Phys. Rev. Lett. 113, 151302.
\bibitem[Gnedin \& Primack (2004)]{Gnedin} Gnedin, O. Y. \& Primack, J. R., 2004, Phys. Rev. Lett. 93, 061302.
\bibitem[Gondolo \& Silk (1999)]{Gondolo} Gondolo, P. \& Silk, J., 1999, Phys. Rev. Lett. 83, 1719.
\bibitem[Hannuksela, Ng \& Li (2020)]{Hannuksela} Hannuksela O. A., Ng K. C. Y. \& Li T. G. F., 2020, Phys. Rev. D 102, 103022.
\bibitem[John, Leane \& Linden (2023)]{John} John I., Leane R. K. \& Linden T., 2023, arXiv:2311.16228.
\bibitem[Kar et al. (2023)]{Kar} Kar A., Kim H., Kim S. P. \& Scopel S., 2023, arXiv:2311.16539.
\bibitem[Kavanagh et al. (2020)]{Kavanagh} Kavanagh, B. J., Nichols, D. A., Bertone, G. \& Gaggero, D., 2020, Phys. Rev. D 102, 083006.
\bibitem[Komossa et al. (2023)]{Komossa} Komossa S. {\it et al.}, 2023, Mon. Not. R. Astron. Soc. 522, L84.
\bibitem[Lacroix \& Silk (2018)]{Lacroix} Lacroix T. \& Silk J., 2018, Astrophys. J. 853, L16.
\bibitem[Laine et al. (2020)]{Laine} Laine S. {\it et al.}, 2020, Astrophys. J. 894, L1.
\bibitem[Li et al. (2022)]{Li} Li, G.-L., Tang, Y. \& Wu Y.-L., 2022, Science China Physics, Mechanics \& Astronomy 65, 100412.
\bibitem[Maggiore (2007)]{Maggiore} Maggiore, M., 2007,  Gravitational Waves: Volume 1: Theory and Experiments (Oxford University Press, New York).
\bibitem[Martinez (2023)]{Martinez} Martinez, D., 2023, Astrophys. Sp. Sci. 368, 45.
\bibitem[McConnell \& Ma (2013)]{McConnell} McConnell N. J. \& Ma C.-P., 2013, Astrophys. J. 764, 184.
\bibitem[Merritt (2004a)]{Merritt} Merritt, D., 2004a, Single and binary black holes and their influence on nuclear structure, in L. Ho (ed.). Coevolution of black holes and galaxies. Carnegie Observatories Astrophysics Series (Cambridge University Press, pp.263-275)(astro-ph/0301257).
\bibitem[Merritt (2004b)]{Merritt2} Merritt, D., 2004b, Phys. Rev. Lett. 92, 201304.
\bibitem[Merritt (2013)]{Merritt3} Merritt, D., 2013, Dynamics and evolution of Galactic nuclei (Princeton University Press, Princeton).
\bibitem[Merritt et al. (2002)]{Merritt4} Merritt, D., Milosavljevic, M., Verde, L. \& Jimenez, R., 2002, Phys. Rev. Lett. 88, 191301.
\bibitem[Mukherjee et al. (2023)]{Mukherjee} Mukherjee D., Holgado A. M., Ogiya G. \& Trac H., 2023, arXiv:2312.02275.
\bibitem[Nampalliwar et al. (2021)]{Nampalliwar} Nampalliwar, S., Saurahb K., Jusufi, K., Wu, Q., Jamil, M. \& Salucci, P., 2021, Astrophys. J. 916, 116.
\bibitem[Navarro et al. (1996)]{Navarro} Navarro, J. F., Frenk, C. S. \& White, S. D. M., 1996, Astrophys. J. 462, 563.
\bibitem[Peters \& Mathews (1963)]{Peters} Peters P. C. \& Mathews J., 1963, Phys. Rev. 131, 435.
\bibitem[Planck Collaboration (2020)]{Planck} Planck Collaboration, 2020, Astron. Astrophys. 641, A6.
\bibitem[Qunbar \& Stone (2023)]{Qunbar} Qunbar I. \& Stone N. C., 2023, arXiv:2304.13062.
\bibitem[Sadeghian et al. (2013)]{Sadeghian} Sadeghian, L., Ferrer, Francesc \& Will, C. M., 2013, Phys. Rev. D 88, 063522.
\bibitem[Shapiro \& Shelton (2016)]{Shapiro} Shapiro, S. L. \& Shelton, J., 2016, Phys. Rev. D 93, 123510.
\bibitem[Shen et al. (2024)]{Shen} Shen Z.-Q., Yuan G.-W., Jiang C.-Z., Tsai Y.-L. S., Yuan Q. \& Fan Y.-Z., 2024, Mon. Not. R. Astron. Soc. 527, 3196.
\bibitem[Sillanp\"a\"a et al. (1988)]{Sillanpaa} Sillanp\"a\"a A., Haarala S., Valtonen M. J., Sundelius B. \& Byrd G. G., 1988, Astrophys. J. 325, 628.
\bibitem[Sofue (2015)]{Sofue} Sofue Y., 2015, Publ. Astron. Soc. Jpn. 67, 75.
\bibitem[Tang et al. (2021)]{Tang} Tang M., Xu Z. \& Wang J., 2021, Chin. Phys. C, 45, 015110.
\bibitem[Titarchuk, Seifina \& Shrader (2023)]{Titarchuk} Titarchuk L., Seifina E. \& Shrader C., 2023, Astron. Astrophys. 671, A159.
\bibitem[Valtonen \& Lehto (1997)]{Valtonen} Valtonen M. J. \& Lehto H. J., 1997, Astrophys. J. 481, L5.
\bibitem[Valtonen et al. (2008)]{Valtonen4} Valtonen M. J. {\it et al.}, 2008, Nature 452, 851.
\bibitem[Valtonen et al. (2010)]{Valtonen2} Valtonen M. J. {\it et al.}, 2010, Celestial Mech. Dyn. Astron. 106, 235.
\bibitem[Valtonen et al. (2023)]{Valtonen3} Valtonen M. J. {\it et al.}, 2023, Mon. Not. R. Astron. Soc. 521, 6143.
\bibitem[Weisberg \& Taylor (2005)]{Weisberg} Weisberg J. M. \& Taylor J. H., 2005, Binary Radio Pulsars ASP Conference Series vol. 328 (astro-ph/0407149).
\bibitem[Xu et al. (2021)]{Xu} Xu W. {\it et al.}, 2021, Astrophys. J. 922, 162.
\bibitem[Young (1980)]{Young} Young P., 1980, Astrophys. J. 242, 1232.
\bibitem[Yue \& Cao (2019)]{Yue} Yue X.-J. \& Cao Z., 2019, Phys. Rev. D 100, 043013.
\bibitem[Zhao, Haehnelt \& Rees (2002)]{Zhao2} Zhao H., Haehnelt M. G. \& Rees M. J., 2002, New Astron. 7, 385.
\bibitem[Zhao et al. (2023)]{Zhao} Zhao Y., Sun B., Lin K. \& Cao Z., 2023, Phys. Rev. D 108, 024070.
\bibitem[Zwick \& Mayer (2023)]{Zwick} Zwick L. \& Mayer L., 2023, Mon. Not. R. Astron. Soc. 526, 2754. 
\end{thebibliography}
\end{document}